\documentclass[10pt,a4paper,prl,twocolumn]{revtex4}
\usepackage{amsmath}
\usepackage{amsfonts}
\usepackage{amssymb}
\usepackage{graphicx}
\usepackage{epsfig}
\begin{document}
\title{Granular segregation under vertical tapping}

\author{Massimo Pica Ciamarra}\email[]{picaciamarra@na.infn.it}
\author{Maria Domenica De Vizia}
\author{Annalisa Fierro}
\author{Marco Tarzia}
\author{Antonio Coniglio}
\author{Mario Nicodemi}

\affiliation{Dip. di Scienze Fisiche, Universit\'a di Napoli 
`Federico II', Coherentia-CNR, INFN, CRdC AMRA, Napoli, Italy}

\date{\today}

\begin{abstract}
We present extensive Molecular Dynamics simulations on species segregation 
in a granular mixture subject to vertical taps. 
We discuss how grain properties, e.g., size, density, friction, 
as well as, shaking properties, e.g., amplitude and frequency, 
affect such a phenomenon. 
Both Brazil Nut Effect (larger particles on the top, BN) and the 
Reverse Brazil Nut Effect (larger particles on the bottom, RBN) are found 
and we derive the system comprehensive ``segregation diagram''
and the BN to RBN crossover line. 
We also discuss the role of friction and show that particles which 
differ only for their frictional properties segregate in states 
depending on the tapping acceleration and frequency. 
\end{abstract}
\maketitle

Granular materials are systems of many particles interacting via short
ranged repulsive and dissipative forces, both normal and tangential to
the surface of contact. They are characterized by an energy scale $mgd$ (of a
grain of mass $m$ and linear size $d$ in the gravitational filed $g$)
which is many orders of magnitue larger than the thermal energy
$k_BT$, and are thus named ``non thermal'' systems. 
These characteristics make difficult the understanding of the
large variety of counterintuitive phenomena granular materials
exhibit, which are of great interest both for their industrial
relevance, and for the theoretical challenges posed to physicist and
engineers. 

Particularly the phenomenon of size segregation under vertical vibrations~\cite{rev_segr},
which we consider here, has emerged as a real conundrum.
Contrary to intuition, an originally disordered mixture when subject
to vertical vibrations tends to order: 
large particles typically rise to the top, as small particles percolate
into their voids during shaking~\cite{rev_segr,Rosato87,Jullien,Duran93} 
or move to the bottom due to convection
mechanisms~\cite{Knight93,Cooke,Gallas}, giving rise to  
the so called ``Brazil Nut Effect" (BN).  
Differences in particle density also affect size separation  (see
ref.s in \cite{Moebius}) and reverse-BN (RBN), with small grains
above, can be observed too~\cite{Shinbrot98,Hong01}. 
The picture where grain sizes and weights are the parameters
explaining segregation is found, however, to be too simple 
\cite{Shinbrot98,Huerta04,Edwards89,Hong01,Levine03,depletion,Jenkins02,Trujillo03,Annalisa0203,Marco04,Jullien92,Both02}
and a full scenario is still missing.

\begin{figure}[t!!]
\includegraphics*[scale=0.32]{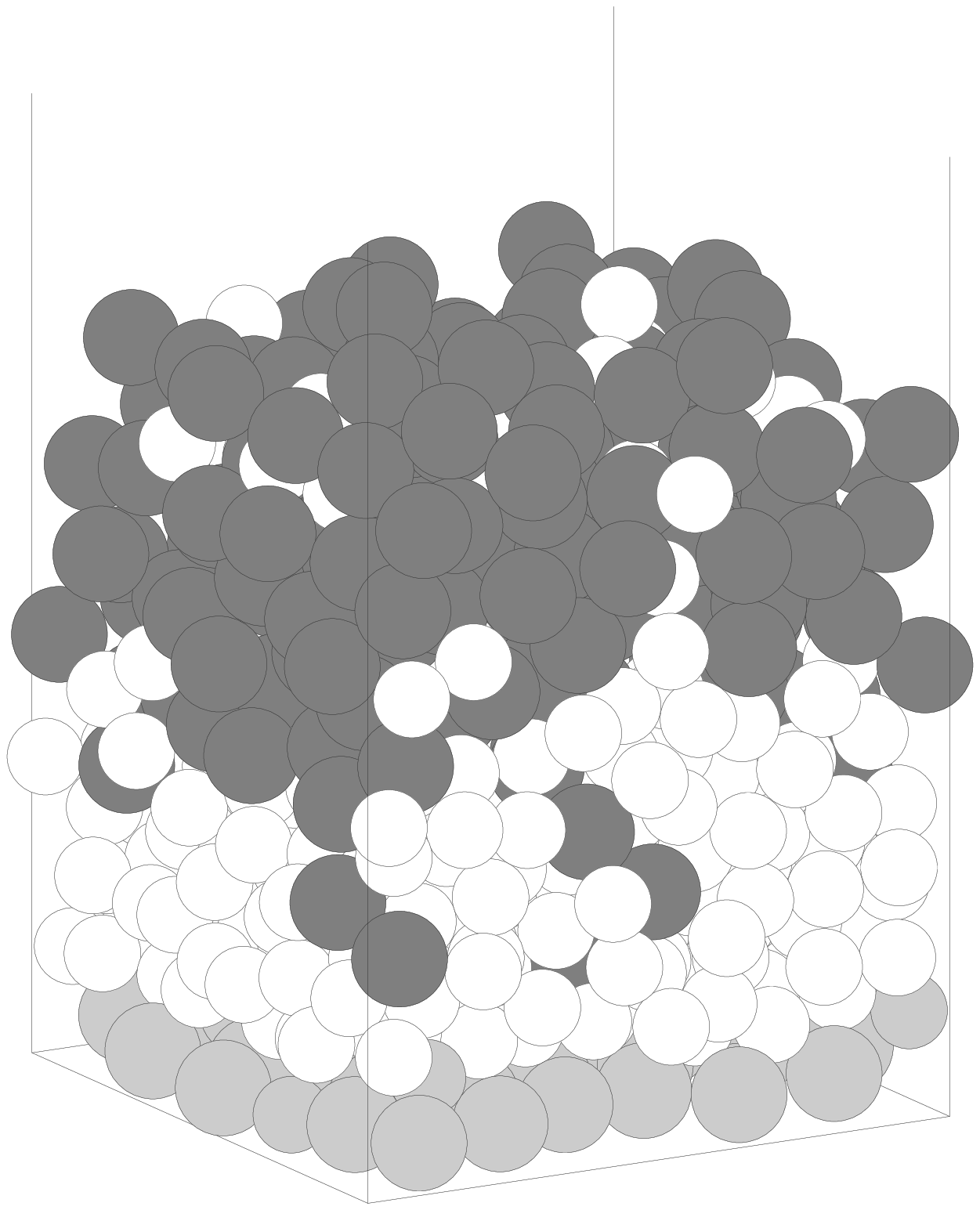}
\includegraphics*[scale=0.32]{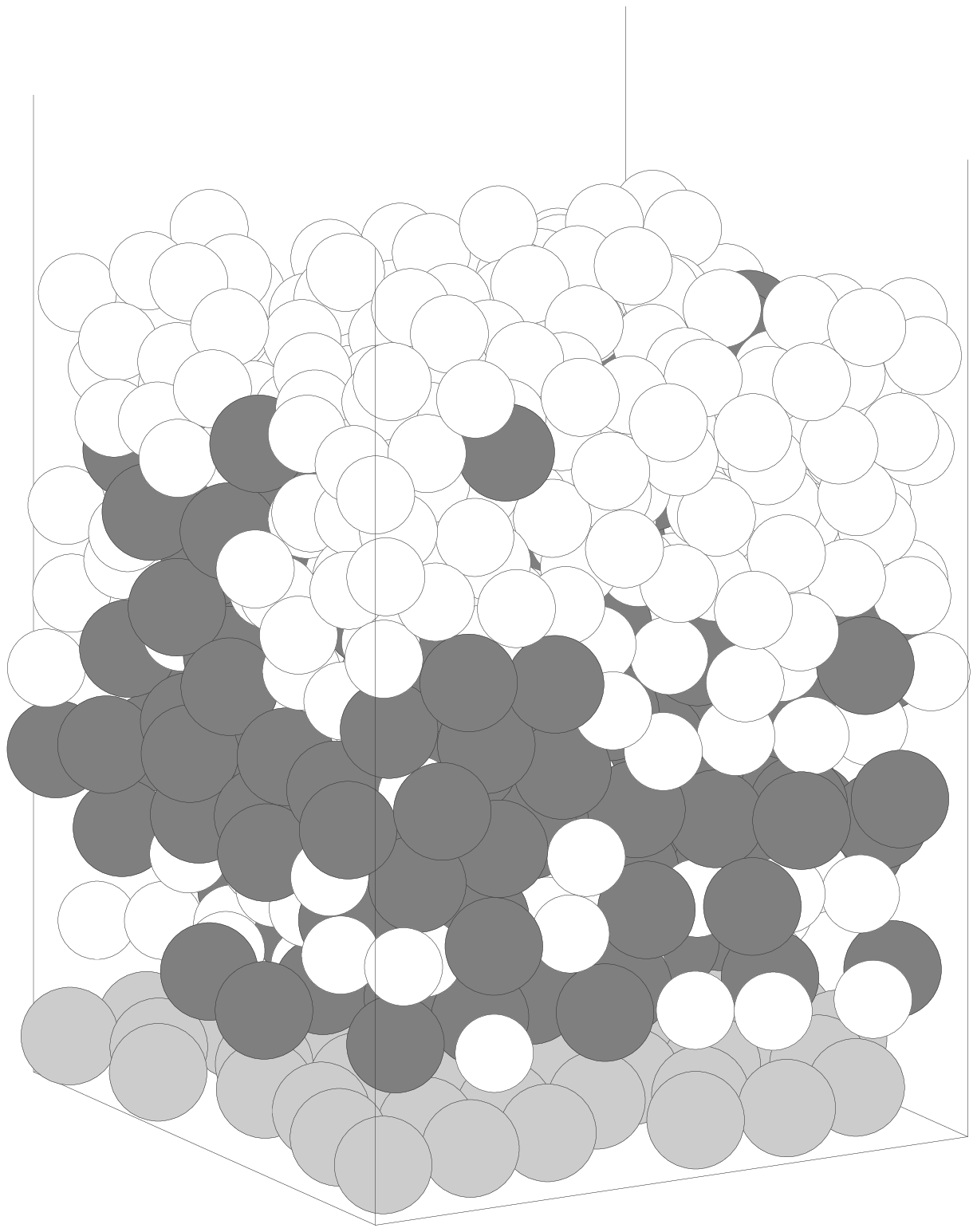}
\caption{\label{fig-plot} We show a mixture of 
$N_l = 240$ large particles of diameter $D_l= 1$ cm and density 
$\rho_l = 1.9$ g~cm$^{-3}$ (dark gray particles) and 
$N_s = 360$ small particles of diameter $D_s = 0.8$ cm and density 
$\rho_s = 1.27$ g~cm$^{-3}$ (white particles). It is contained in a box 
(whose base is made of other immobile grains, light gray particles, see text)
and is subject to vertical taps with normalized amplitude $\Gamma$. 
The pictures show two configurations at rest attained at stationarity: 
interestingly, when shaken with $\Gamma = 1$ (left) the system goes into a BN
configuration and when $\Gamma = 3$ (right) it goes in a RBN
configuration. 
}
\end{figure}

In correspondence with some existing experiments 
\cite{Hsiau97,Canul98,Breu03}, here we consider segregation phenomena 
in Molecular Dynamics simulations of tap dynamics: 
grains confined in a box are shaken and after each shake fully 
dissipate their kinetic energy before being shaken again. 
A picture of our model system is given in the left panel of
Fig.~\ref{fig-plot} showing the final BN configuration reached by 
an initially disordered mixture shaken with an amplitude 
$\Gamma = A \omega^2/g=1$ 
(where $\omega$ is the shake frequency, $A$ its amplitude and $g$ 
gravity acceleration, see below). An example of the role of the external
drive on segregation can be appreciated by comparison with the right 
panel of Fig.~\ref{fig-plot} showing the final RBN
configuration reached by the same mixture when shaken at $\Gamma=3$. 

We show below how grain properties, e.g., size, density, friction, 
as well as the external forcing, e.g., shaking amplitude and frequency, 
affect the process and derive for the first time a comprehensive non
trivial ``segregation diagram''. The richness of such a
diagram is not captured by current theoretical approaches
\cite{Hong01,Both02,Jenkins02} and calls for new theoretical and
experimental investigations.

\textit{Simulations --}
We make soft-core molecular dynamics simulations of a system of $N_l =
240$ large grains of diameter $D_l = 1$ cm and density $\rho_l = 1.9$
g cm$^{-3}$, and  $N_s$ small grains with diameter $D_s$ and the
density $\rho_s$. We vary $D_s$ and $\rho_s$ and chose the number 
$N_s$ in such a way that the two species occupy a comparable volume, 
$N_l D_l^3 \simeq N_s D_s^3$. 
The particles are enclosed in a box with a square basis of
side length $L = 7$ cm (see Fig.\ref{fig-plot}) 
with periodic boundary conditions in the horizontal directions, so
that convection is avoided.
In order to prevent crystallization some particles are randomly glued
on the container basis (in such a way that no further particle can
touch the bottom of the container). 

Two grains interact when in contact via a normal and a tangential
force. The normal force is given by the so-called linear
spring-dashpot model, while the tangential interaction is implemented
by keeping track of the elastic shear displacement throughout the
lifetime of a contact \cite{Si01}. 
The model and the values of its parameters have
been described in \cite{modello}, with 
the value of the viscous coefficient of the normal interaction
such that the restitution coefficient is $e= 0.8$. 
In most of our simulation the static friction coefficients are equal 
for the two species, $\mu_{ll} = \mu_{ls} = \mu_{ss} = 0.4$ 
($\mu_{ij}$ is the friction coefficient in the 
interaction between a grain of type $i$ and a grain of type $j$, and 
$\mu_{ij} = \mu_{ji}$), but we will also consider the case $\mu_{ll} \neq
\mu_{ss}$ to investigate the role of friction in the segregation process.

The system, starting from a random configuration, 
is subject to a tap dynamics up to reach a stationary state. 
Each tap consists of one oscillation of the container basis  with
amplitude $A$ and frequency $\omega$, i.e., the bottom of the box
moves with $z(t) = A\cos(2\pi \omega t)$. 
We checked that both $A$ and $\omega$ are important to select the final
segregation state ($\Gamma$ is not the only relevant parameter) 
and consider here the case where $\omega = 30$ Hz and $A$ is varied. 
A tap is followed by a relaxation time where the system comes to rest. 
A grain is considered to be at rest if its kinetic energy 
becomes smaller than
$10^{-5}$ mgd, where $1$ mgd is the energy required to rise it of
a distance equal to its diameter. 
All measures are taken when the system is at rest and in the
stationary part of the tap dynamics. 
Actually, it is known that for small values of $\Gamma$ the system dynamics 
has strong ``glassy'' features \cite{capri,Bideau_rev} and thus the
states attained can be very far from stationarity. Here we are away
from that region. 


\begin{figure}[t]\vspace{-1cm}
\centerline{\hspace{-2cm}
\epsfig{figure=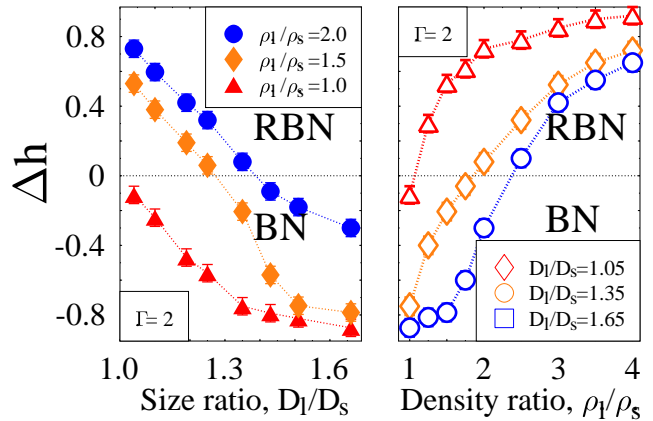,width=8.5cm,angle=-90}}
\vspace{-1.5cm}
\caption{\label{fig-dh-size}
(color on-line) The left panel shows the segregation parameter $\Delta h$ 
as a function of the diameter ratio $D_l/D_s$ of the mixture
components, for $\Gamma =2$ and $\rho_l/\rho_s = 1,1.5,2$. 
By increasing  $D_l/D_s$ the system crosses from RBN configurations
(i.e., $\Delta h>0$) to BN (i.e., $\Delta h<0$). 
The right panel shows $\Delta h$ as a function of the density ratio 
$\rho_l/\rho_s$ for $\Gamma =2$ and $D_l/D_s = 1.05, 1.35, 1.65$.
By increasing $\rho_l/\rho_s$ the system moves 
from RBN to BN, as the crossover point (where $\Delta h=0$) turns out
to depend on the size ratio $D_l/D_s$.
}
\end{figure}

The degree of separation of the binary mixture in the stationary state 
is quantified by the usual vertical segregation parameter
\begin{equation}
\Delta h = 2\frac{h_s-h_l}{h_s+h_l},
\end{equation}
where $h_{p}$ is the average height of particles of species $p = l,s$ 
($h_{p} = 1/N_{p}\sum_{i=1}^{N_p} z_i$, 
here $z_i$ is the height of particle $i$ 
with respect to the container basis at rest). We
prepare the system in a random initial state characterized by $\Delta
h \simeq 0$ via a Monte Carlo procedure. When subject to a tap
dynamics the mixture evolves and the segregation parameter changes
until a stationary state is reached.

\textit{Results--}
We first describe the dependence of the segregation parameter, $\Delta h$, 
on the diameter ratio $D_l/D_s$ and on the density ratio $\rho_l/\rho_s$
of the two components for $\Gamma = 2$.
Fig.~\ref{fig-dh-size} shows $\Delta h$ as a function of $D_l/D_s$ for
different values of the density ratio (left panel) and $\Delta h$ as a
function of $\rho_l/\rho_s$ for different values of the size ratio
(right panel). 
As expected, when the diameter ratio grows BN states are
favoured with respect to RBN states, even though such an effect is
mitigated by increasing the density ratio $\rho_l/\rho_s$ which shifts
the BN to RBN crossover to higher values of $D_l/D_s$.

\begin{figure}[t!]\vspace{-1cm}
\centerline{\hspace{-2cm}
\epsfig{figure=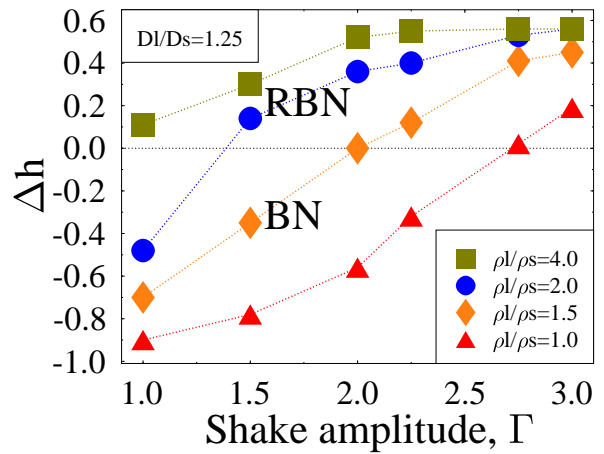,width=9cm,angle=-90}}
\vspace{-1.5cm}
\caption{\label{fig-dh-gamma}
(color on-line) $\Delta h$ is plotted as a
function of the adimensional vibrational acceleration $\Gamma$, 
in a mixture where $D_l/D_s=1.25$, for the shown values 
of the density ratio $\rho_l/\rho_s$. 
As $\Gamma$ increases the system moves from a BN to a RBN configuration.
}
\end{figure}
Such a size and density ratio dependence might seem to result from two 
simple competing effects. The first one is a ``percolation'' 
effect~\cite{Rosato87} according to which it is easier for the smaller
particles to percolate through the voids between larger grains and, thus, 
reach the bottom of the container. 
The percolation effect becomes stronger as the size ratio between the 
components increases, and therefore should describe the size
dependence found in Fig.~\ref{fig-dh-size} (left panel). 
The second effect is buoyancy, according to which the system tends 
to minimize gravitational energy, and therefore the species with higher 
mass density is pushed to the bottom of the container. This should 
describe the density dependence of Fig.~\ref{fig-dh-size} (right panel). 

We find, however, that the properties of the external forcing have an
essential role in selecting the final segregation state and the
overall scenario appears to be richer: percolation/buoyancy effects
changes with the intensity of vibration $\Gamma$ (see Fig.\ref{fig-plot}). 
In Fig.~\ref{fig-dh-gamma} we plot the dependence of the
segregation parameter, $\Delta h$, on $\Gamma$ for given values of 
$D_l/D_s$ and $\rho_l/\rho_s$: unexpectedly, 
a stronger shaking enhances RBN, i.e., as $\Gamma$ increases 
$\Delta h$ increases too. A similar qualitative result was observed also 
in experiments with a continuous shaking dynamics~\cite{Breu03}. 

\begin{figure}[t!!]
\includegraphics*[scale=0.42]{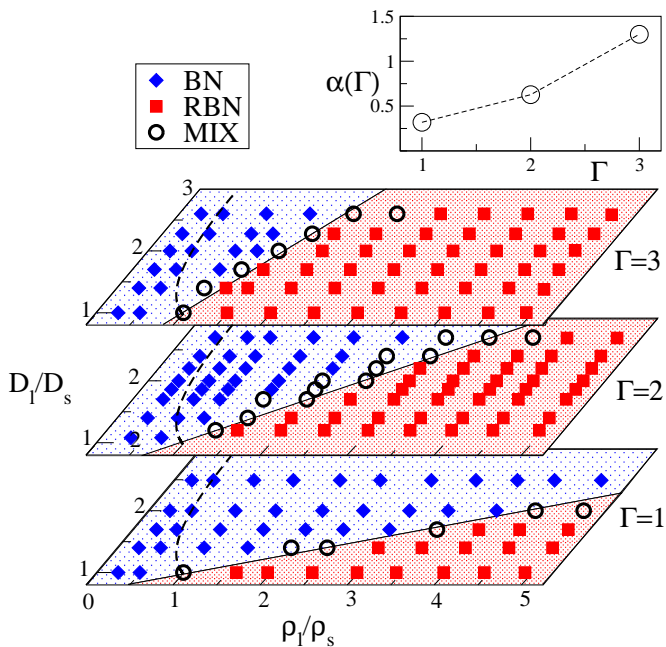}
\caption{\label{fig-fase-dr}
(color on-line) The ``segregation diagram'' of the mixture 
in the ($\rho_l/\rho_s$, $D_l/D_s$, $\Gamma$) space. 
The plot shows the regimes where the reverse (RBN) and the
usual Brazil nut effect (BN) occur. Empty circles are the points where 
$\Delta h$ is zero, within $10\%$, and named ``mix'' in the caption. 
The solid line separating the areas is given by
Eq.~(\ref{eq-sqrt}). The dashed line is the 
crossover line proposed in Ref.\cite{Hong01} 
Inset: dependence on the adimensional acceleration $\Gamma$ 
of the coefficient $\alpha$ of Eq.~(\ref{eq-sqrt}).
}
\end{figure}

Fig.~\ref{fig-fase-dr} summarizes these findings in a ``segregation 
diagram'' in the ($\rho_l/\rho_s$, $D_l/D_s$, $\Gamma$) space. 
As expected, the BN effect is favored when $D_l/D_s$ is large and RBN
when $\rho_l/\rho_s$ grows. The BN to RBN crossover region is
dependent on $\Gamma$: we approximate 
the BN to RBN crossover line $D_l/D_s=f(\rho_l/\rho_s,\Gamma)$ around
$\rho_l/\rho_s\simeq 1$  with a linear function (continuous line in
the figure):
\begin{equation}
\label{eq-sqrt}
\frac{D_l}{D_s} \simeq 1+\alpha(\Gamma) \left(\frac{\rho_l}{\rho_s}-1\right)
\end{equation}
where the angular coefficient $\alpha(\Gamma)$, shown in the inset of 
Fig.\ref{fig-fase-dr}, grows monotonically with $\Gamma$. 
Since $\alpha(\Gamma) > 0$, the present results, corresponding to 
grains with equal
friction properties (see below for a different case), point out that 
RBN configurations can be found only if $\rho_l/\rho_s > 1$: 
i.e., by changing $D_l/D_s$ there is no way to find RBN when
$\rho_l/\rho_s < 1$. 
In this perspective our simulations may explain why in the
``original'' Brazil Nut Effect observed during the transportation 
of nuts of different size (but otherwise similar) the larger ones are
systematically found to rise to the surface. 
The diagram of Fig.~\ref{fig-fase-dr} appears to be in good agreement
with the general features
of known experiments as those of Ref.~\cite{Breu03} (even though it is
still unclear whether the phenomena of segregation under tapping, here
considered, and under continuous shaking, as in \cite{Breu03}, are
qualitatively similar) and can help clarifying experimental results. 

In Fig.~\ref{fig-fase-dr} we also plot the BN to RBN crossover line 
found by the theory of Ref.\cite{Hong01} (dashed line): 
such a theory, approximating the granular mixture under vibration as a 
thermal system, predicts the right qualitative behavior as a function 
of $\rho_l/\rho_s$, but it doesn't capture the right $D_l/D_s$ dependence. 
More elaborated models assuming lack of equipartition between the
species, as in  kinetic theories \cite{Jenkins02} or some simulations
\cite{Trujillo03}, or the existence of more than one configurational
temperature, as in a Statistical Mechanics theory of the mixture 
\cite{Annalisa0203,Marco04}, may be able to improve on this aspect. 

\begin{figure}[t!]
\includegraphics[scale = 0.32]{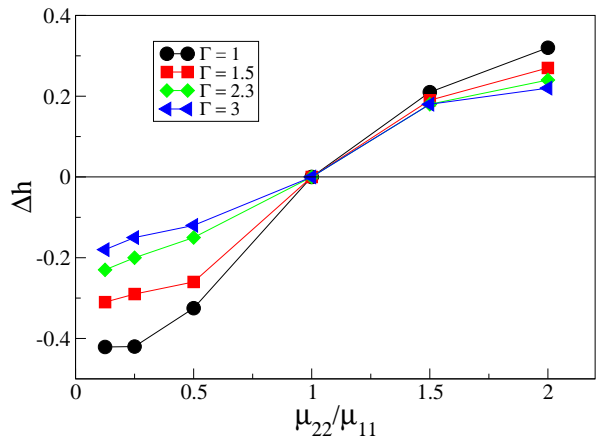}
\caption{\label{fig-friction}
(color on-line) 
The segregation parameter $\Delta h$ is shown as a function of the
ratio between the static friction coefficient $\mu_{22}/\mu_{11}$ 
of a mixture of grains which differ only for their frictional
properties. Particles with higher friction coefficient are always
found on the top of the container, althought the degree of segregation
depends on $\Gamma$. 
}
\end{figure}


An other parameter relevant to segregation is friction
\cite{Levine03,Behringer} which we now consider in order to extend the
diagram of Fig.~\ref{fig-fase-dr}. We take grains with equal sizes
and weights, but different friction coefficients; this is an interesting
situation difficult to be experimentally accessed as real grains
which differ in frictional properties usually also differ in other 
properties (such as mass, Young modulus, etc...). 
We study a mixture of $N_1 = N_2 = 300$ grains of
diameter $D_1 = D_2 = 1$ cm and density $\rho_1 = \rho_2 = 1.9$ g
cm$^{-3}$ with friction coefficients $\mu_{11} = 0.4$, 
$\mu_{22}\in\{0.05,0.01,0.2,0.4,0.6,0.8\}$, 
and $\mu_{12} = \min\left(\mu_{11},\mu_{12}\right)$, as in~\cite{Levine03}. 
As the two components only differ for their friction, the 
segregation parameter is now defined as $\Delta h = 2{(h_2-h_1)}/{(h_2+h_1)}$.
This mixture indeed segregates: Fig.~\ref{fig-friction} shows that 
the species with higher friction coefficient always rise to the top, 
as the degree of segregation depends on the shaking intensity. 
This can be explained by considering that grains with smaller
friction can  more easily percolate to the bottom of the container.


\textit{Conclusions --}
Our Molecular Dynamics simulations are not affected by the presence of air,
humidity and (due to the periodic boundary conditions) convection, and
should be therefore considered as an ideal, even though comprehensive, 
experiment, well robust to changes in the MD model \cite{Si01}. 
We found that 
both grain properties, such as diameters, densities and friction, and 
external driving properties, such as amplitude and frequency of shaking, 
are important to select the system final segregation state. 
We determined the ``segregation diagram'' in the three
parameters space ($D_l/D_s$, $\rho_l/\rho_s$, $\Gamma$) and 
derived the BN to RBN crossover line $D_l/D_s=f(\rho_l/\rho_s,\Gamma)$. 
In particular, in our model system, a mixture of grains 
only differing in sizes always segregates in a BN configuration, 
explaining why in the original ``Brazil nut'' problem large grains
always seat at the top. 
We also discussed how segregation is influenced by grains friction
by showing, for instance, that in a mixture of particles 
differing only for their surface friction by increasing $\Gamma$ the
smoother grains tend to rise to the top, a result easy to be
experimentally checked. 
As our results are in agreement with known experiments and can
help their clearer interpretation, our comprehensive ``segregation
diagram'' is not explained by current theories and necessitates
further theoretical and experimental investigations. 

\begin{acknowledgments}
We thank M. Schr\"oter for useful discussions. 
Work supported by EU Network MRTN-CT-2003-504712, MIUR-PRIN 2004,
MIUR-FIRB 2001, CRdC-AMRA.

\end{acknowledgments}

\begin{thebibliography}{40}
\bibitem{modello}
We use the model `L3' described in detail in Ref.~\cite{Si01}, 
neglecting the rotational motion of the spheres.
\end{thebibliography}

\end{document}